\documentclass[submission, Phys, hidelnks]{SciPost}
\usepackage{physics}
\usepackage{graphicx}
\usepackage[ruled, vlined]{algorithm2e}
\usepackage{amsmath, bm}
\usepackage{dsfont}
\usepackage{listings}
\usepackage{color}
\usepackage{xcolor}
\DeclareFixedFont{\ttb}{T1}{txtt}{bx}{n}{10}
\DeclareFixedFont{\ttm}{T1}{txtt}{m}{n}{10}
\definecolor{deepblue}{rgb}{0,0,0.7}
\definecolor{deepred}{rgb}{0.6,0,0}
\definecolor{deepgreen}{rgb}{0,0.5,0}
\newcommand\pythonstyle{\lstset{
    language=Python,
    keepspaces=true,
    basicstyle=\ttfamily\small,
    tabsize=3,
    keywordstyle=\color{keywordcolour},
    frame=single,
    keywordstyle=\color{deepblue},
    emph={SigmaX, SigmaZ, __init__, RenyiEntropy, swap},
    emphstyle=\color{deepred},
    stringstyle=\color{deepgreen},
    showstringspaces=false,
    commentstyle=\color{deepgreen}\ttfamily
}}
\lstnewenvironment{python}[1][]{
    \pythonstyle{}
    \lstset{#1}
}
{}
\newcommand\pythoninline[1]{{\pythonstyle\lstinline!#1!}}

\newcommand{\x}{\bm{\mathrm{x}}}
\SetKwComment{Comment}{$\triangleright$\ }{}

\begin{document}

\begin{center}{\Large \textbf{
    QuCumber: wavefunction reconstruction with neural networks
}}\end{center}

\begin{center}
    Matthew~J.~S.~Beach\textsuperscript{1,2},
    Isaac~De~Vlugt\textsuperscript{2},
    Anna~Golubeva\textsuperscript{1,2},
    Patrick~Huembeli\textsuperscript{1,3},
    Bohdan~Kulchytskyy\textsuperscript{1,2},
    Xiuzhe~Luo\textsuperscript{2},
    Roger~G.~Melko\textsuperscript{1,2*},
    Ejaaz~Merali\textsuperscript{2},
    Giacomo~Torlai\textsuperscript{1,2,4}
\end{center}

\begin{center}
    ${\bf 1}$ Perimeter Institute for Theoretical Physics, Waterloo,
    Ontario N2L 2Y5, Canada
    \\
    \mbox{${\bf 2}$ Department of Physics and Astronomy, University of Waterloo,
    Ontario N2L 3G1, Canada}
    \\
    ${\bf 3}$ ICFO-Institut de Ciencies Fotoniques, Barcelona Institute of Science and Technology,
    08860 Castelldefels (Barcelona), Spain
    \\
    ${\bf 4}$ Center for Computational Quantum Physics, Flatiron Institute,
    162 5th Avenue, New York, NY 10010, USA\\

    * \href{mailto:rgmelko@uwaterloo.ca}{rgmelko@uwaterloo.ca}
\end{center}
\begin{center}
\today
\end{center}
\section*{Abstract}
{\bf
As we enter a new era of quantum technology, it is increasingly important to
develop methods to aid in the accurate preparation of quantum states
for a variety of materials, matter, and devices.
Computational techniques can be used to reconstruct a state from data,
however the growing number of qubits demands ongoing algorithmic advances
in order to keep pace with experiments.
In this paper, we present an open-source software package called QuCumber that
uses machine learning to reconstruct a quantum state consistent
with a set of projective measurements. QuCumber uses a restricted
Boltzmann machine to efficiently represent the quantum wavefunction for a large
number of qubits.
New measurements can be generated from the machine to
obtain physical observables not easily accessible from the original data.
}

\noindent\rule{\textwidth}{1pt}
\tableofcontents\thispagestyle{fancy}
\noindent\rule{\textwidth}{1pt}
\vspace{-1cm}

\section{Introduction}
Current advances in fabricating quantum technologies, as well as in reliable
control of synthetic quantum matter, are leading to a new era of quantum
hardware where highly pure quantum states are routinely prepared in laboratories.
With the growing number of controlled quantum degrees of freedom, such as
superconducting qubits, trapped ions, and ultracold
atoms~\cite{kandala_hardware-efficient_2017,moll_quantum_2018,bernien_probing_2017,zhang_observation_2017}, reliable and scalable
classical algorithms are required for the analysis and verification of
experimentally prepared quantum states. Efficient algorithms can aid in
extracting physical observables otherwise inaccessible from experimental
measurements, as well as in identifying sources of noise to provide direct
feedback for improving experimental hardware. However, traditional approaches
for reconstructing unknown quantum states from a set of measurements, such as
quantum state tomography, often suffer the exponential overhead that is typical
of quantum many-body systems.

Recently, an alternative path to quantum state reconstruction was put forward,
based on modern machine learning (ML)
techniques~\cite{torlai_learning_2016,torlai_neural-network_2018,torlai_latent_2018,carrasquilla_reconstructing_2018,lennon_efficiently_2018,kim_mixed_2018}.
The most common approach relies on a powerful generative model called a
\textit{restricted Boltzmann machine} (RBM)~\cite{smolensky_information_1986},
a stochastic neural network with two layers of binary units.
A visible layer $\bm{v}$ describes the physical degrees of freedom, while a
hidden layer $\bm{h}$ is used to capture high-order correlations between the
visible units. Given a set of neural network parameters
 $\bm{\lambda}$, the RBM defines a probabilistic model described by the
 parametric distribution $p_{\bm{\lambda}}(\bm{v})$.
RBMs have been widely used in the ML community for the pre-training of deep
neural networks~\cite{hinton_training_2002}, for compressing high-dimensional
data into lower-dimensional representations~\cite{hinton_reducing_2006}, and
more~\cite{lecun_deep_2015}.
More recently, RBMs have been adopted by the physics community in the context
of representing both classical and quantum many-body
states~\cite{CarleoTroyer2017Science,carleo_constructing_2018}.
They are currently being investigated for their representational
power~\cite{gao_efficient_2017,choo_symmetries_2018,glasser_neural-network_2018},
their relationship with tensor networks and the renormalization
group~\cite{mehta_exact_2014,koch-janusz_mutual_2018,iso_scale-invariant_2018,lenggenhager_optimal_2018,chen_equivalence_2018},
and in other contexts in quantum many-body physics~\cite{PhysRevB.96.205152,weinstein_neural_2018,RBM_stabilizer}.

In this post, we present QuCumber: a \textit{quantum calculator used for
many-body eigenstate reconstruction}. QuCumber is an open-source Python
package that implements neural-network quantum state reconstruction of
many-body wavefunctions from projective measurement data.
Examples of data to which QuCumber could be applied might
be magnetic spin projections, orbital occupation number,
polarization of photons, or the logical state of qubits.
Given a training set of such measurements, QuCumber discovers the most likely
compatible quantum state by finding the optimal set of
parameters $\bm{\lambda}$ of an RBM.\@ A properly trained RBM is an
approximation of the unknown quantum state underlying the data.
It can be used to calculate various physical observables of interest, including
measurements that may not be possible in the original experiment.

This post is organized as follows.
In Section~\ref{sec:positive}, we introduce the reconstruction technique for
the case where all coefficients of the wavefunction are real and positive.
We discuss the required format for input data, as well as training of the RBM
and the reconstruction of typical observables. In Section~\ref{sec:complex},
we consider the more general case of a complex-valued wavefunction.
We illustrate a general strategy to extract the phase structure from data by
performing appropriate unitary rotations on the state before measurements.
We then demonstrate a practical reconstruction of an entangled state of two
qubits.
Note, the detailed theory underlying the reconstruction methods used by QuCumber
can be found in the original references \cite{torlai_learning_2016,torlai_neural-network_2018} and a
recent review \cite{RBMreview}.
A list of useful terms and equations can be found at the end of this
post in the Glossary.

\section{Positive wavefunctions}\label{sec:positive}
We begin by presenting the application of QuCumber to reconstruct many-body
quantum states described by wavefunctions $\ket{\Psi}$ with positive
coefficients $\Psi(\x)=\braket{\x}{\Psi} \ge 0$,
where $\ket{\x}=\ket{ \mathrm{x}_1,\dots,\mathrm{x}_N }$ is a reference basis
for the Hilbert space of $N$ quantum degrees of freedom.
The neural-network quantum state reconstruction requires raw data
$\mathcal{D}=(\x_1,\x_2,\dots)$ generated through projective measurements of
the state $\ket{ \Psi}$ in the reference basis.
These measurements adhere to the probability distribution given by the Born
rule, $P(\x)=|\Psi(\x)|^2$. Since the wavefunction is strictly positive, the
quantum state is completely characterized by the measurement distribution,
i.e.~$\Psi(\x)=\sqrt{P(\x)}$.

The positivity of the wavefunction allows a simple and natural connection
between quantum states and classical probabilistic models. QuCumber employs the
probability distribution $p_{\bm{\lambda}}(\x)$ of an RBM (see Eq.~\ref{Eq:marginal_distribution} of the Glossary)
to approximate the
distribution $P(\x)$ underlying the measurement data. Using contrastive
divergence (CD)~\cite{hinton_training_2002}, QuCumber trains the RBM to discover
an optimal set of parameters ${\bm \lambda}$ that minimize the Kullback-Leibler (KL) divergence
between the two distributions (see Eq.~\ref{Eq.KLdiv}). Upon successful training
($p_{\bm{\lambda}}(\x)\sim P(\x)$), we obtain an approximate representation of
the target quantum state,
\begin{equation}\label{wfpd}
    \psi_{\bm{\lambda}}(\x) \equiv \sqrt{p_{\bm{\lambda}}(\x)}
    \simeq\Psi(\x)\:.
\end{equation}
Note, the precise mathematical form of the marginal distribution $p_{\bm{\lambda}}(\x)$ defined in terms of an effective energy
over the parameters of the RBM is defined in the Glossary.

In the following, we demonstrate the application of QuCumber for the reconstruction of the
ground-state wavefunction of the one-dimensional transverse-field Ising model
(TFIM). The Hamiltonian is
\begin{equation}
    \hat{H} = -J\sum_i \hat{\sigma}^z_i \hat{\sigma}^z_{i+1} - h \sum_i\hat{\sigma}^x_i\:, \label{TFIM}
\end{equation}
where $\hat{\sigma}^{x/z}_i$ are spin-1/2 Pauli operators acting on site $i$,
and we assume open boundary conditions. For this example, we consider a chain
with $N=10$ spins at the quantum critical point $J=h=1$.

\subsection{Setup}\label{subsec:example}
Given the small size of the system, the ground state $\ket{ \Psi}$ can be found
with exact diagonalization. The training dataset $\mathcal{D}$ is generated by
sampling the distribution $P(\bm{\sigma}^z)=|\Psi(\bm{\sigma}^z)|^2$, obtaining
a sequence of $N_S=10^5$ independent spin projections in the reference basis
$\x = \bm{\sigma}^z$.\footnote{
        The training dataset can be downloaded from
        \href{https://github.com/PIQuIL/QuCumber/blob/master/examples/Tutorial1_TrainPosRealWaveFunction/tfim1d_data.txt}{\texttt{https://github.com/PIQuIL/QuCumber/blob/master/\\examples/Tutorial1\_TrainPosRealWaveFunction/tfim1d\_data.txt}}
    }
Each data point in $\mathcal{D}$ consists of an array $\bm{\sigma}^z_j=(\sigma^z_1,\dots,\sigma^z_N)$ with shape \verb|(N,)| and should be passed to QuCumber as a numpy array or torch tensor. For example, $\bm{\sigma}^z_j=$ \verb|np.array([1,0,1,1,0,1,0,0,0,1])|, where we use $\sigma_j^z=0,1$ to represent a spin-down and spin-up state respectively. Therefore, the entire input data set is contained in an array with shape \verb|(N_S, N)|.

Aside from the training data, QuCumber also allows us to import an exact
wavefunction. This can be useful for monitoring the quality of the
reconstruction during training. In our example, we evaluate the fidelity between
the reconstructed state $\psi_{\bm{\lambda}}(\x)$ and the exact wavefunction
$\Psi(\x)$. The training dataset, \verb|train_data|, and the exact ground
state, \verb|true_psi|, are loaded with the data loading utility as follows:
\begin{python}
import qucumber.utils.data as data
train_path = "tfim1d_data.txt"
psi_path = "tfim1d_psi.txt"
train_data, true_psi = data.load_data(train_path, psi_path)
\end{python}
If \verb|psi_path| is not provided, QuCumber will load only the training data.

Next, we initialize an RBM quantum state $\psi_{\bm{\lambda}}(\bf{x})$ with
random weights and zero biases using the constructor \verb|PositiveWaveFunction|:
\begin{python}
from qucumber.nn_states import PositiveWaveFunction
state = PositiveWaveFunction(num_visible=10, num_hidden=10)
\end{python}
The number of visible units (\verb|num_visible|) must be equal to the number
of physical spins $N$, while the number of hidden units (\verb|num_hidden|) can
 be adjusted to systematically increase the representational power of the RBM.\@

The quality of the reconstruction will depend on the structure underlying the
specific quantum state and the ratio of visible to hidden units,
$\alpha = \verb|num_hidden|/\verb|num_visible|$.
In practice, we find that $\alpha = 1$ often leads to good
approximations of positive wavefunctions~\cite{torlai_neural-network_2018}.
However, in the general case, the value of $\alpha$ required for a given
wavefunction should be explored and adjusted by the user.

\subsection{Training}
Once an appropriate representation of the quantum state has been defined,
QuCumber trains the RBM through the function \verb|PositiveWaveFunction.fit|.
Several input parameters need to be provided aside from the training dataset
(\verb|train_data|). These include the number of training iterations
(\verb|epochs|), the number of samples used for the positive/negative phase of
CD (\verb|pos_batch_size|/\verb|neg_batch_size|), the learning rate (\verb|lr|)
and the number of sampling steps in the negative phase of CD (\verb|k|). The
last argument (\verb|callbacks|) allows the user to pass a set of additional
functions to be evaluated during training.

As an example of a callback, we show the \verb|MetricEvaluator|, which
evaluates a function \verb|log_every| epochs during training.
Given the small system size and the knowledge of the true ground state, we can
evaluate the fidelity between the RBM state and the true ground-state
wavefunction (\verb|true_psi|).
Similarly, we can calculate the KL divergence between the RBM distribution
$p_{\bm{\lambda}}(\x)$, and the data distribution $P(\x)$, which should
approach zero for a properly trained RBM.\@
For the current example, we monitor the fidelity and KL divergence (defined in
\verb|qucumber.utils.training_statistics|):
\begin{python}
from qucumber.callbacks import MetricEvaluator
import qucumber.utils.training_statistics as ts
log_every = 10
space = state.generate_hilbert_space(10)
callbacks = [
    MetricEvaluator(
        log_every,
        {"Fidelity": ts.fidelity, "KL": ts.KL},
        target_psi=true_psi,
        space=space,
        verbose=True
    )
]
\end{python}
With \verb|verbose=True|, the program will print the epoch number and all
callbacks every \verb|log_every| epochs.
For the current example, we monitor the fidelity and KL divergence (see Glossary).
Note that the KL divergence is only tractable for small systems.
The \verb|MetricEvaluator| will compute the KL exactly when provided with a list of all states in the Hilbert space. For convenience these can be generated with \\
\verb|space = state.generate_hilbert_space(10)|.

\begin{figure}[hbt]
    \centering{}
    \includegraphics[width=\columnwidth, trim={0 15 0 0}, clip]{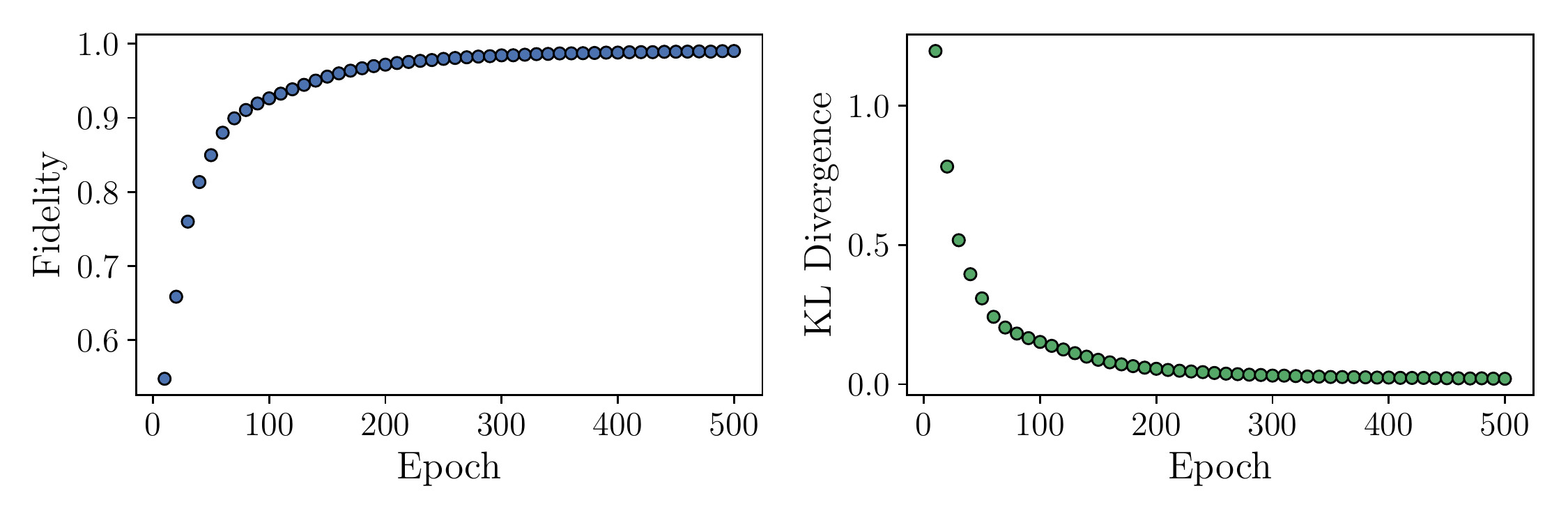}
    \caption{
        The fidelity (left) and the KL divergence (right) during training for
        the reconstruction of the ground state of the one-dimensional TFIM.\@
    }\label{fig:KL}
\end{figure}

Now that the metrics to monitor during
training have been chosen, we can invoke the optimization with the \verb|fit|
function of \verb|PositiveWaveFunction|.
\begin{python}
state.fit(
    train_data,
    epochs=500,
    pos_batch_size=100,
    neg_batch_size=100,
    lr=0.01,
    k=5,
    callbacks=callbacks,
)
\end{python}
Figure~\ref{fig:KL} shows the convergence of the fidelity and KL divergence
during training. The convergence time will, in general, depend on the choice
of hyperparameters.
Finally, the network parameters $\bm{\lambda}$, together with the \verb|MetricEvaluator|'s data,
can be saved (or loaded) to a file:
\begin{python}
state.save(
    "filename.pt",
    metadata={
        "fidelity": callbacks[0].Fidelity,
        "KL": callbacks[0].KL
    },
)
state.load("filename.pt")
\end{python}

With this we have demonstrated the most basic aspects of QuCumber regarding
training a model and verifying its accuracy. We note that in this example the
evaluation utilized the knowledge of the exact ground state and the calculation
of the KL divergence, which we reemphasize is tractable only for small system sizes.  However,
we point out that QuCumber is capable of carrying out the reconstruction of much
larger systems. In such cases, users must rely on other estimators to evaluate
the training, such as expectation values of physical observables (magnetization,
energy, etc). In the following, we show how to compute diagonal and
off-diagonal observables in QuCumber.

\subsection{Reconstruction of physical observables}\label{Sec:Sampling_a-Trained_RBM}
In this section, we discuss how to calculate the average value of a generic
physical observable $\hat{\mathcal{O}}$ from a trained RBM.\@
We start with the case of observables that are diagonal in the reference basis
where the RBM was trained. We then discuss the more general cases of
off-diagonal observables and entanglement entropies.

\subsubsection{Diagonal observables}
We begin by considering an observable with only diagonal matrix elements,
$\bra{ \bm{\sigma} } \hat{\mathcal{O}} \ket{ \bm{\sigma}^{\prime} }=\mathcal{O}_{\bm{\sigma}}\delta_{\bm{\sigma\sigma}^\prime}$
where for convenience we denote the reference basis ${\bf x}=\bm{\sigma}^z$ as
$\bm{\sigma}$ unless otherwise stated.
The expectation value of $\hat{\mathcal{O}}$ is given by
\begin{equation}
    \ev*{\hat{\mathcal{O}}} = \frac{1}{\sum_{\bm{\sigma}} |\psi_{\bm{\lambda}}(\bm{\sigma})|^2}
    \sum_{\bm{\sigma}} \mathcal{O}_{\bm{\sigma}}|\psi_{\bm{\lambda}}(\bm{\sigma})|^2\:.
\end{equation}
The expectation value can be approximated by a Monte Carlo estimator,
\begin{equation}
    \ev*{\hat{\mathcal{O}}} \approx \frac{1}{N_{\rm MC}} \sum_{k=1}^{N_{\rm MC}} \mathcal{O}_{\bm{\sigma}_k}\:,
\end{equation}
where the spin configurations $\bm{\sigma}_k$ are sampled from the RBM
distribution $p_{\bm{\lambda}}(\bm{\sigma})$. This process is particularly
efficient given the bipartite structure of the network which allows the use
of block Gibbs sampling.

A simple example for the TFIM is the average longitudinal magnetization per
spin, $\ev{\hat\sigma^z} = \sum_j\ev*{\hat{\sigma}^z_j}/N$,
which can be calculated directly on the spin configuration sampled by the RBM
(i.e.,~the state of the visible layer). The visible samples are obtained with
the \verb|sample| function of the RBM state object:
\begin{python}
samples = state.sample(num_samples=1000, k=10)
\end{python}
which takes the total number of samples (\verb|num_samples|) and the number
of block Gibbs steps (\verb|k|) as input. Once these samples are obtained,
the magnetization can be calculated simply as
\begin{python}
magnetization = samples.mul(2.0).sub(1.0).mean()
\end{python}
where we converted the binary samples of the RBM back into $\pm 1$ spins before
taking the mean.

\subsubsection{Off-diagonal observables}

We turn now to the case of off-diagonal observables, where the expectation
value assumes the following form
\begin{equation}
    \ev*{\hat{\mathcal{O}}} = \frac{1}{\sum_{\bm{\sigma}} |\psi_{\bm{\lambda}}(\bm{\sigma})|^2}
    \sum_{\bm{\sigma\sigma}^\prime} \psi_{\bm{\lambda}}^*(\bm{\sigma})
    \psi_{\bm{\lambda}}(\bm{\sigma}^\prime)\mathcal{O}_{\bm{\sigma\sigma}^\prime}\:.
\end{equation}
This expression can once again be approximated with a Monte Carlo estimator
\begin{equation}
    \ev*{\hat{\mathcal{O}}} \approx \frac{1}{N_{\rm MC}} \sum_{k=1}^{N_{\rm MC}} \mathcal{O}^{[L]}_{\bm{\sigma}_k}
\end{equation}
of the so-called \emph{local estimator} of the observable:
\begin{equation}
    \mathcal{O}^{[L]}_{\bm{\sigma}_k}=\sum_{\bm{\sigma}^\prime}\frac{\psi_{\bm{\lambda}}(\bm{\sigma}^\prime)}{\psi_{\bm{\lambda}}(\bm{\sigma}_k)} \mathcal{O}_{\bm{\sigma}_k\bm{\sigma}^\prime}\:.
\end{equation}
As long as the matrix representation $\mathcal{O}_{\bm{\sigma\sigma}^\prime}$
is sufficiently sparse in the reference basis, the summation can be evaluated
efficiently.

\begin{figure}[t]
    \centering{}
    \includegraphics[width=\columnwidth, trim={0 15 0 0}, clip]{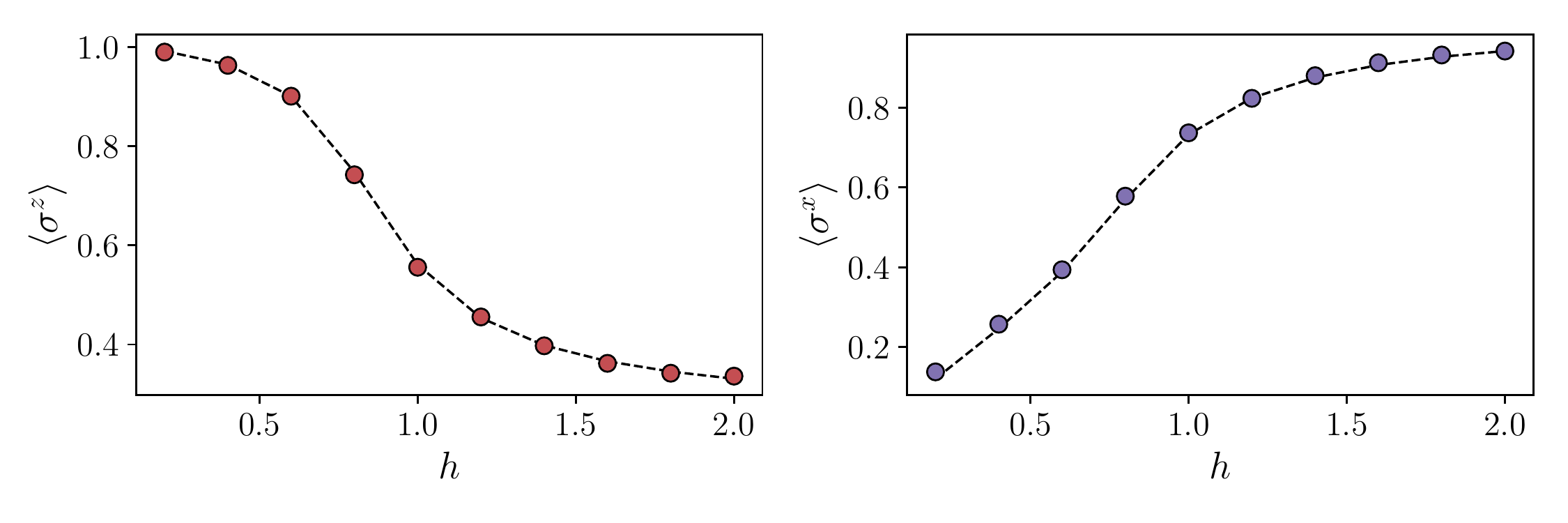}
    \caption{
        Reconstruction of the magnetic observables for the TFIM chain with
        $N=10$ spins. We show the average longitudinal (left) and transverse
        (right) magnetization per site obtained by sampling from a trained RBM.\@
        The dashed line denotes the results from exact diagonalization.
    }\label{tfim_magn}
\end{figure}

As an example, we consider the specific case of the transverse magnetization
for the $j$-th spin, $\ev*{\hat{\sigma}^x_j}$, with matrix elements
\begin{equation}
    \mel{\bm{\sigma}}{\hat{\sigma}^x_j}{\bm{\sigma}^{\prime}}=\delta_{\sigma_j^\prime,1-\sigma_j}\prod_{i\ne j}\delta_{\sigma_i^\prime,\sigma_j}\:.
\end{equation}
Therefore, the expectation values reduces to the Monte Carlo average of the
local observable
\begin{equation}
    {(\sigma^x_j)}^{[L]}=\frac{\psi_{\bm{\lambda}}(\sigma_1,\dots,1-\sigma_j,\dots,\sigma_N)}
    {\psi_{\bm{\lambda}}(\sigma_1,\dots,\sigma_j,\dots,\sigma_N)}
\:.
\end{equation}
evaluated on spin configurations $\bm{\sigma}_k$ sampled from the RBM
distribution $p_{\bm{\lambda}}(\bm{\sigma})$.

QuCumber provides an interface for sampling off-diagonal observables in the
\verb|ObservableBase| class. Thorough examples are available in the tutorial
section in the \href{https://qucumber.readthedocs.io/en/stable/}{documentation}.\!\!
\footnote{The observables tutorial is available at
    \href{https://qucumber.readthedocs.io/en/stable/\_examples/Tutorial3\_DataGeneration\_CalculateObservables/tutorial\_sampling\_observables.html
}{\texttt{https://qucumber.readthedocs.io/en/stable/\_examples\\/Tutorial3\_DataGeneration\_CalculateObservables/tutorial\_sampling\_observables.html}}
}
As an example, $\sigma^x$ can be written as an observable class with
\begin{python}
import torch
from qucumber.utils import cplx
from qucumber.observables import ObservableBase
class SigmaX(ObservableBase):

    def apply(self, nn_state, samples):
        psi = nn_state.psi(samples)
        psi_ratio_sum = torch.zeros_like(psi)

        for i in range(samples.shape[-1]):  # sum over spin sites
            flip_spin(i, samples)  # flip the spin at site i
            # add ratio psi_(-i) / psi to the running sum
            psi_flip = nn_state.psi(samples)
            psi_ratio = cplx.elementwise_division(psi_flip, psi)
            psi_ratio_sum.add_(psi_ratio)
            flip_spin(i, samples)  # flip it back

        # take real part and divide by number of spins
        return psi_ratio_sum[0].div_(samples.shape[-1])
\end{python}
The value of the observable can be estimated from a set of samples with:
\begin{python}
SigmaX().statistics_from_samples(state, samples)
\end{python}
which produces a dictionary containing the mean, variance, and standard error of
the observable. Similarly, the user can define other observables like the energy.

The reconstruction of two magnetic observables for the TFIM is shown in
Fig.~\ref{tfim_magn}, where a different RBM was trained for each value of
the transverse field $h$. In the left plot, we show the average longitudinal
magnetization per site, which can be calculated directly from the
configurations sampled by the RBM.\@ In the right plot, we show the off-diagonal
observable of transverse magnetization. In both cases, QuCumber successfully
discovers an optimal set of parameters $\bm{\lambda}$ that accurately
approximate the ground-state wavefunction underlying the data.

\subsubsection{Entanglement entropy}\label{sec:swap}

A quantity of significant interest in quantum many-body systems is the degree of
entanglement between a subregion $A$ and its complement $\bar{A}$.
Numerically, measurement of bipartite entanglement entropy is commonly accessed
through the computation of the second R\'enyi entropy $S_2 = - \ln {\rm Tr}(\rho_A^2)$.
When one has access to a pure state wavefunction $\psi_{\bm{\lambda}}(\x)$,
R\'enyi entropies can be calculated as an expectation value of the ``Swap'' operator~\cite{Swap},
\begin{equation}\label{Eq:renyi_entropy}
S_2 = - \ln \left\langle{   \widehat{\textrm{Swap}}_A  }\right\rangle.
\end{equation}
It is essentially an off-diagonal observable that acts on an extended product space consisting of
two independent copies of the wavefunction,
$\psi_{\bm{\lambda}}(\x) \otimes \psi_{\bm{\lambda}}(\x)$, referred to as
``replicas''.
As the name suggests, the action of the Swap operator is to swap the spin
configurations in region $A$ between the replicas,
\begin{equation}
  \widehat{\textrm{Swap}}_A |\bm{\sigma}_A, \bm{\sigma}_{\bar A}\rangle_1 \otimes  |\bm{\sigma}^{\prime}_A, \bm{\sigma}^{\prime}_{\bar A}\rangle_2 = |\bm{\sigma}^{\prime}_A, \bm\sigma_{\bar A}\rangle_1 \otimes  |\bm\sigma_A, \bm\sigma^{\prime}_{\bar A}\rangle_2 .
\end{equation}
Here the subcript of the ket indicates the replica index, while the two labels
inside a ket, such as ${\bm \sigma}_A, {\bm \sigma}_{\bar A}$, describe the spins
configurations within the subregion and its complement.

In QuCumber, the Swap operator is implemented as a routine within the
\verb|entanglement| observable unit,
\begin{python}
def swap(s1, s2, A):
    _s = s1[:, A].clone()
    s1[:, A] = s2[:, A]
    s2[:, A] = _s
    return s1, s2
\end{python}
where \verb|s1| and \verb|s2| are batches of samples produced from each replica,
and \verb|A| is a list containing the indices of the sites in subregion $A$.
While ideally those samples should be entirely independent, in order to save
computational costs, QuCumber just splits a given batch into two equal parts and
treats them as if they were independent samples. This is implemented within the
\verb|SWAP| observable,
\begin{python}
class SWAP(ObservableBase):
    def __init__(self, A):
        self.A = A

    def apply(self, nn_state, samples):
        _ns = samples.shape[0] // 2
        samples1 = samples[:_ns, :]
        samples2 = samples[_ns : _ns * 2, :]

        psi_ket1 = nn_state.psi(samples1)
        psi_ket2 = nn_state.psi(samples2)

        psi_ket = cplx.elementwise_mult(psi_ket1, psi_ket2)
        psi_ket_star = cplx.conjugate(psi_ket)

        samples1_, samples2_ = swap(samples1, samples2, self.A)
        psi_bra1 = nn_state.psi(samples1_)
        psi_bra2 = nn_state.psi(samples2_)

        psi_bra = cplx.elementwise_mult(psi_bra1, psi_bra2)
        psi_bra_star = cplx.conjugate(psi_bra)
        return cplx.real(
            cplx.elementwise_division(psi_bra_star, psi_ket_star)
        )
\end{python}
Note the similarity in the implementation to that for the transverse
magnetization observable from the last section, once the amplitude of a sample
is substituted with the product of amplitudes drawn from each replica.

Using this observable, we can estimate the R\'enyi entropy of the region containing
the first 5 sites in the chain using Eq.~\ref{Eq:renyi_entropy},
\begin{python}
A = [0, 1, 2, 3, 4]
swap_ = SWAP(A)
swap_stats = swap_.statistics_from_samples(state, samples)
S_2 = -np.log(swap_stats["mean"])
\end{python}

\begin{figure}[htb]
    \centering
    \includegraphics[]{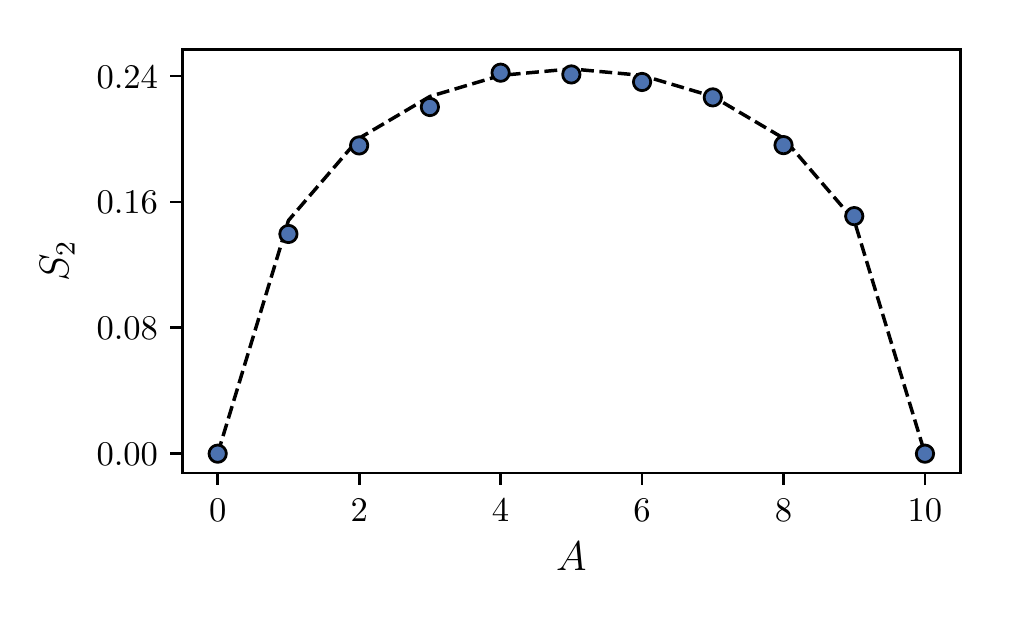}
    \caption{
        The second R\'enyi entropy for the TFIM chain with
        $N=10$ spins.
        The number of sites in the entangled bipartition $A$
        is indicated by the horizontal axis.
        The markers indicate values obtained through the
        ``Swap'' operator applied  to the samples from a trained RBM.\@
        The dashed line denotes the result from exact diagonalization.
    }\label{ee_learn}
\end{figure}

We apply this measurement procedure to a TFIM chain with results shown in
Fig.~\ref{ee_learn}. As was the case with the magnetization observables, the trained
RBM gives a good approximation to the second R\'enyi entropy for different subregion
$A$ sizes. Being a basis-independent observable, this constitutes a useful test on
the ability of QuCumber to capture the full wavefunction from the information contained
in a single-basis dataset for TFIM.

\section{Complex wavefunctions}\label{sec:complex}
For positive wavefunctions, the probability distribution underlying the outcomes
of projective measurements in the reference basis  contains all possible
information about the unknown quantum state.
However, in the more general case of a wavefunction with a non-trivial sign or
phase structure, this is not the case.
In this section, we consider a target quantum state where the wavefunction
coefficients in the reference basis can be
complex-valued, $\Psi(\bm{\sigma})=\Phi(\bm{\sigma})e^{i\theta(\bm{\sigma})}$.
We continue to choose the reference basis as $\bm{\sigma} = \bm{\sigma}^z$.
We first need to generalize the RBM representation of the quantum state to
capture a generic complex wavefunction. To this end, we introduce an additional RBM
with marginalized distribution $p_{\bm{\mu}}(\bm{\sigma})$ parameterized by a
new set of network weights and biases $\bm{\mu}$.
We use this to define the quantum state as:
\begin{equation}
    \psi_{\bm{\lambda} \bm{\mu}} (\bm{\sigma})= \sqrt{p_{\bm{\lambda}} (\bm{\sigma})} e^{i \phi_{\bm{\mu}} (\bm{\sigma})/2}
\end{equation}
where $\phi_{\bm{\mu}}(\bm{\sigma}) = \log (p_{\bm{\mu}} (\bm{\sigma}))$~\cite{torlai_neural-network_2018}.
In this case, the reconstruction requires a different type of measurement
setting. It is easy to see that projective measurements in the reference basis
do not convey any information on the phases $\theta(\bm{\sigma})$, since
$P(\bm{\sigma})=|\Psi(\bm{\sigma})|^2=\Phi^2(\bm{\sigma})$.

\begin{figure}[htb]
    \centering
    \includegraphics[width=\columnwidth, trim={0 0 0 40}, clip]{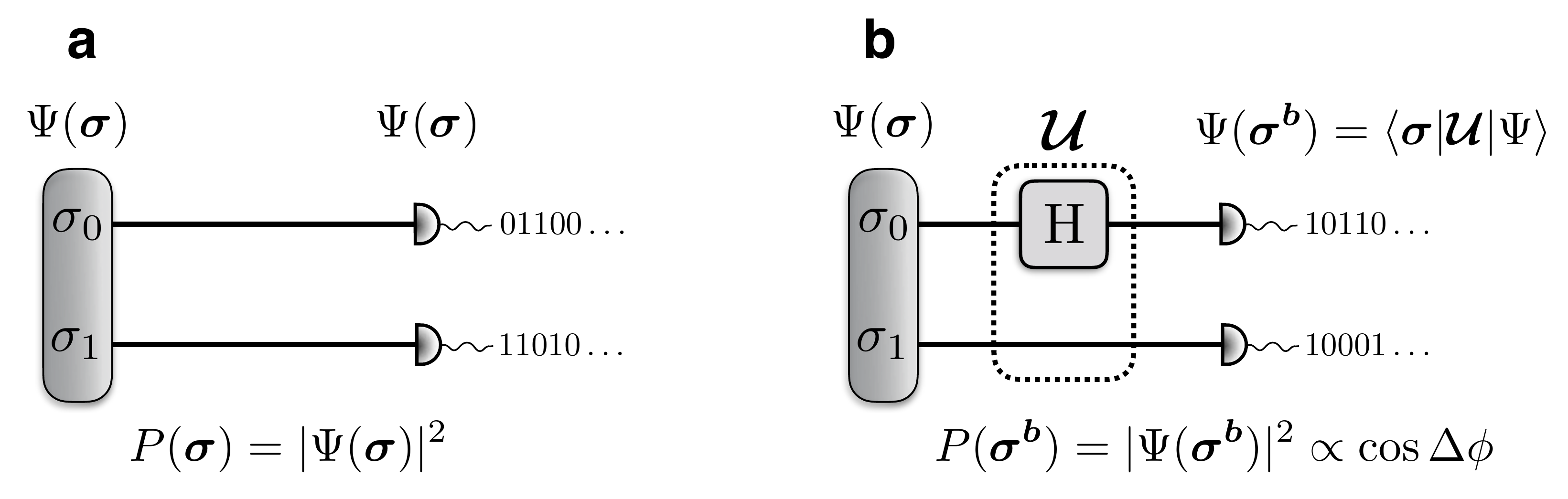}
    \caption{
        Unitary rotations for two qubits. (left) Measurements on the reference
        basis. (right) Measurement in the rotated basis. The unitary rotation
        (the Hadamard gate on qubit $\sigma_0$) is applied after state
        preparation and before the projective measurement.
    }\label{phase_learn}
\end{figure}

The general strategy to learn a phase structure is to apply a unitary
transformation $\bm{\mathcal{U}}$ to the state $\ket{\Psi}$ before the
measurements, such that the resulting measurement distribution
$P^{\:\prime}(\bm{\sigma})=|\Psi^\prime(\bm{\sigma})|^2$ of the rotated state
$\Psi^\prime(\bm{\sigma})=\bra{ \bm{\sigma} } \:\bm{\mathcal{U}}\:\ket{ \Psi}$
contains fingerprints of the phases $\theta(\bm{\sigma})$
(Fig.~\ref{phase_learn}). In general, different rotations must be independently
applied to gain full information on the phase structure. We make the assumption
of a tensor product structure of the rotations,
$\bm{\mathcal{U}}=\bigotimes_{j=1}^N\hat{\mathcal{U}}_j$. This is equivalent to
a local change of basis from $\ket{ \bm{\sigma}}$ to
$\lbrace|\bm{\sigma}^{\bm{b}}\rangle=|\sigma_1^{b_1},\dots,\sigma_N^{b_N}\rangle\rbrace$,
where the vector $\bm{b}$ identifies the local basis $b_j$ for each site $j$.
The target wavefunction in the new basis is given by
\begin{equation}
\begin{split}
    \Psi(\bm{\sigma}^{\bm{b}})
    &=\langle \bm{\sigma}^{\bm{b}}|\Psi\rangle=\sum_{\bm{\sigma}}\langle \bm{\sigma}^{\bm{b}}|\bm{\sigma}\rangle\langle\bm{\sigma}|\Psi\rangle\\
    &=\sum_{\bm{\sigma}}\bm{\mathcal{U}}(\bm{\sigma}^{\bm{b}},\bm{\sigma})\Psi(\bm{\sigma})\:,
\end{split}
\end{equation}
and the resulting measurement distribution is
\begin{equation}
    P_{\bm{b}}(\bm{\sigma}^{\bm{b}})=\bigg|\sum_{\bm{\sigma}}\bm{\mathcal{U}}(\bm{\sigma}^{\bm{b}},\bm{\sigma})\Psi(\bm{\sigma})\bigg|^2\:.
\end{equation}

To clarify the procedure, let us consider the simple example of a quantum state
of two qubits:
\begin{equation}
    |\Psi\rangle=\sum_{\sigma_0,\sigma_1}\Phi_{\sigma_0\sigma_1}e^{i\theta_{\sigma_0\sigma_1}}|\sigma_0\sigma_1\rangle\:,
\end{equation}
and rotation $\bm{\mathcal{U}}=\hat{\mathrm{H}}_0\otimes\hat{\mathcal{I}}_1$,
where $\hat{\mathcal{I}}$ is the identity operator and
\begin{equation}
    \hat{\mathrm{H}}=\frac{1}{\sqrt{2}}\begin{bmatrix}1 & 1\\
1 & -1
\end{bmatrix}
\end{equation}
is called the {\it Hadamard gate}. This transformation is equivalent to
rotating the qubit $\sigma_0$ from the reference $\sigma_0^z$ basis the the
$\sigma_0^x$ basis. A straightforward calculation leads to the following
probability distribution of the projective measurement in the new basis
$|\sigma_0^x,\sigma_1\rangle$:
\begin{equation}
    P_{\bm{b}}(\sigma_0^x,\sigma_1)=\frac{\Phi_{0\sigma_1}^2+\Phi_{1\sigma_1}^2}{4}+\frac{1-2\sigma_0^x}{2}\Phi_{0\sigma_1}\Phi_{1\sigma_1}\cos(\Delta\theta)\:,
\end{equation}
where $\Delta\theta=\theta_{0\sigma_1}-\theta_{1\sigma_1}$. Therefore, the
statistics collected by measuring in this basis implicitly contains partial
information on the phases. To obtain the full phases structure, additional
transformations are required, one example being the rotation from the reference
basis to the $\sigma^y_j$ local basis, realized by
the elementary gate
\begin{equation}
        \hat{\mathrm{K}}=\frac{1}{\sqrt{2}}\begin{bmatrix}1 & -i\\
    1 & i
\end{bmatrix}\:.
\end{equation}

\subsection{Setup}
We now proceed to use QuCumber to reconstruct a complex-valued wavefunction.
For simplicity, we restrict ourselves to two qubits and consider the general case
of a quantum state with random amplitudes $\Phi_{\sigma_0\sigma_1}$ and random
phases $\theta_{\sigma_0\sigma_1}$. This example is available in the online tutorial.
\footnote{The tutorial for complex wavefunctions can be found at
    \href{https://qucumber.readthedocs.io/en/stable/\_examples/Tutorial2_TrainComplexWaveFunction/tutorial_qubits.html
    }{\texttt{https://qucumber.readthedocs.io/en/\\stable/\_examples/Tutorial2\_TrainComplexWaveFunction/tutorial\_qubits.html}}
}
We begin by importing the required packages:

\begin{python}
from qucumber.nn_states import ComplexWaveFunction
import qucumber.utils.unitaries as unitaries
import qucumber.utils.cplx as cplx
\end{python}
Since we are dealing with a complex wavefunction, we load the corresponding module
\verb|ComplexWaveFunction| to build the RBM quantum state
$\psi_{\bm{\lambda\mu}}(\bm{\sigma})$. Furthermore, the following additional
utility modules are required: the \verb|utils.cplx| backend for complex algebra,
and the \verb|utils.unitaries| module  which contains a set of elementary local
rotations. By default, the set of unitaries include local rotations to the
$\sigma^x$ and $\sigma^y$ basis, implemented by the $\hat{\mathrm{H}}$ and
$\hat{\mathrm{K}}$ gates respectively.

We continue by loading the data\footnote{
    The training dataset can be downloaded from
    \href{https://github.com/PIQuIL/QuCumber/blob/master/examples/Tutorial2_TrainComplexWaveFunction/}{\texttt{https://github.com/PIQuIL/QuCumber/blob/master/\\examples/Tutorial2\_TrainComplexWaveFunction/}}
} into QuCumber, which is done using the
\verb|load_data| function of the data utility:
\begin{python}
train_path = "qubits_train.txt"
train_bases_path = "qubits_train_bases.txt"
psi_path = "qubits_psi.txt"
bases_path = "qubits_bases.txt"

train_samples, true_psi, train_bases, bases = data.load_data(
  train_path, psi_path, train_bases_path, bases_path
)
\end{python}
As before, we may load the true target wavefunction from \verb|qubits_psi.txt|,
which can be used to calculate the fidelity and KL divergence. In contrast with
the positive case, we now have measurements performed in different bases.
Therefore, the training data consists of an array of qubits projections
$(\sigma_0^{b_0},\sigma_1^{b_1}$) in \verb|qubits_train_samples.txt|, together
with the corresponding bases $(b_0,b_1)$ where the measurement was taken, in
\verb|qubits_train_bases.txt|. Finally, QuCumber loads the set of all the bases
appearing the in training dataset, stored in \verb|qubits_bases.txt|. This is
required to properly configure the various elementary unitary rotations that
need to be applied to the RBM state during the training. For this example, we
generated measurements in the following bases:
\begin{equation}
    (b_0, b_1)=(\mathrm{z},\mathrm{z})\:,\:(\mathrm{x},\mathrm{z})\:,\:(\mathrm{z},\mathrm{x})\:,\:(\mathrm{y},\mathrm{z})\:,\:(\mathrm{z},\mathrm{y})
\end{equation}
Finally, before the training, we initialize the set of unitary rotations and
create the RBM state object. In the case of the provided dataset, the unitaries
are the $\hat{\mathrm{H}}$ and $\hat{\mathrm{K}}$ gates. The required dictionary
can be created with \verb|unitaries.create_dict()|.
By default, when \verb|unitaries.create_dict()| is called, it will contain the
identity, the $\hat{\mathrm{H}}$ gate, and the $\hat{\mathrm{K}}$ gate, with the
keys \verb|Z|, \verb|X|, and \verb|Y|, respectively. It is possible to add
additional gates by specifying them as
\begin{python}
U = torch.tensor([[<re_part>], [<im_part>]], dtype=torch.double)
unitary_dict = unitaries.create_dict(<unitary_name>=U)
\end{python}
where \verb|re_part|, \verb|im_part|, and \verb|unitary_name| are to be
specified by the user.

We then initialize the complex RBM object with
\begin{python}
state = ComplexWaveFunction(
   num_visible=2, num_hidden=2, unitary_dict=unitary_dict
)
\end{python}
The key difference between positive and complex wavefunction reconstruction is
the requirement of additional measurements in different basis. Despite this,
loading the data, initializing models, and training the RBMs are all very
similar to the positive case, as we now discuss.

\subsection{Training}
Like in the case of a positive wavefunction, for the complex case
QuCumber optimizes the network parameters to minimize the KL divergence between
the data and the RBM distribution. When measuring in multiple bases, the
optimization now runs over the set of parameters $(\bm{\lambda},\bm{\mu})$ and
minimizes the sum of KL divergences between the data distribution
$P(\bm{\sigma}^{\bm{b}})$ and the RBM distribution
 $|\psi_{\bm{\lambda\mu}}(\bm{\sigma}^{\bm{b}})|^2$ for each bases $\bm{b}$
 appearing in the training dataset~\cite{torlai_neural-network_2018}.
For example, if a given training sample is measured in the basis
 $(\mathrm{x},\mathrm{z})$, QuCumber applies the appropriate unitary
 rotation $\bm{\mathcal{U}}=\hat{\mathrm{H}}_0\otimes\hat{\mathcal{I}}_1$ to
 the RBM state before collecting the gradient signal.

\begin{figure}[htb]
    \centering{}
    \includegraphics[width=\textwidth, trim={0 15 0 0}, clip]{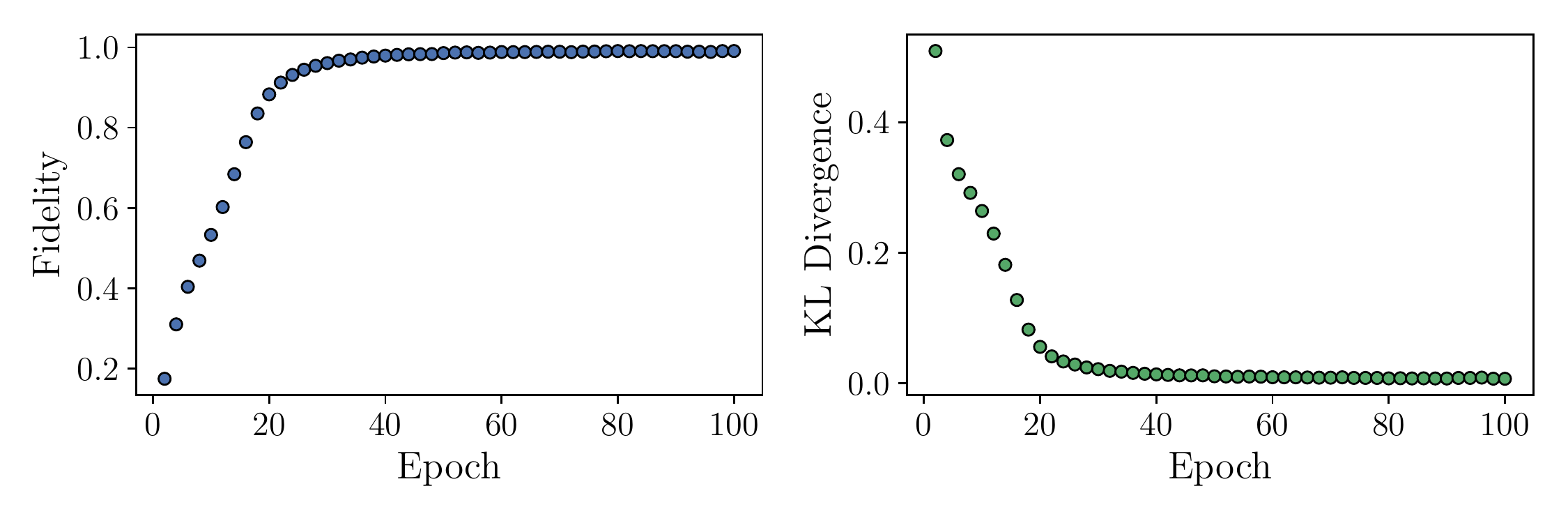}
    \caption{
        Training a complex RBM with QuCumber on random two-qubit data. We
        show the fidelity (left), and KL divergence (right), as a function of
        the training epochs.
    }\label{fig:complex}
\end{figure}

Similar to the case of positive wavefunction, we generate the Hilbert space (to
compute fidelity and KL divergence) and initialize the callbacks
\begin{python}
state.space = nn_state.generate_hilbert_space(2)
callbacks = [
  MetricEvaluator(
    log_every,
    {"Fidelity": ts.fidelity, "KL": ts.KL},
    target_psi=true_psi,
    bases=bases,
    verbose=True,
    space=state.space,
  )
]

\end{python}
The training is carried out by calling the \verb|fit| function of
\verb|ComplexWaveFunction|, given the set of hyperparameters
\begin{python}
state.fit(
    train_samples,
    epochs=100,
    pos_batch_size=10,
    neg_batch_size=10,
    lr=0.05,
    k=5,
    input_bases=train_bases,
    callbacks=callbacks,
)
\end{python}
In Fig.~\ref{fig:complex} we show the total KL divergence and the fidelity with
the true two-qubit state during training.
After successfully training a QuCumber model, we can once again compute
expectation values of physical observables,
as discussed in Section~\ref{Sec:Sampling_a-Trained_RBM}.

\section{Conclusion}

We have introduced the open source software package QuCumber, a quantum
calculator used for many-body eigenstate reconstruction. QuCumber is capable
of taking input data representing projective measurements of a quantum
wavefunction, and reconstructing the wavefunction using a restricted Boltzmann
machine (RBM).

Once properly trained, QuCumber can produce a new set of measurements,
sampled stochastically from the RBM.\@ These samples, generated in the
reference basis, can be used to verify the training of the RBM against the
original data set. More importantly, they can be used to calculate expectation
values of many physical observables. In fact, any expectation value typically
estimated by conventional Monte Carlo methods can be implemented as an
estimator in QuCumber. Such estimators may be inaccessible in the reference
basis, for example. Or, they may be difficult or impossible to implement in the
setup for which the original data was obtained.  This is particularly relevant
for experiments, where it is easy to imagine many possible observables that are
inaccessible, due to fundamental or technical challenges.

Future versions of QuCumber, as well as the next generation of quantum state
reconstruction software, may explore different generative models, such as
variational autoencoders, generative adversarial networks, or recurrent neural networks.
The techniques described in this paper can also be extended to reconstruct mixed states,
via the purification technique described in Reference \cite{torlai_latent_2018}.
In addition, future techniques may include hybridization between machine
learning and other well-established methods in computational quantum many-body
physics, such as variational Monte Carlo and tensor networks \cite{carrasquilla_reconstructing_2018}.

\section*{Acknowledgements}
We acknowledge M. Albergo, G. Carleo, J. Carrasquilla, D. Sehayek, and
L. Hayward Sierens for stimulating discussions.
We thank the Perimeter Institute for Theoretical Physics for the continuing support of PIQuIL.\@

\paragraph{Author contributions}
Authors are listed alphabetically. For an updated record of individual
contributions, consult the repository
at \url{https://github.com/PIQuIL/QuCumber/graphs/contributors}.

\paragraph{Funding information}

This research was supported by the Natural Sciences and Engineering
Research Council of Canada (NSERC), the
Canada Research Chair program, and the Perimeter Institute
for Theoretical Physics. We also gratefully
acknowledge the support of NVIDIA Corporation with
the donation of the Titan Xp GPU used in this work.
Research at Perimeter Institute is supported by the Government
of Canada through Industry Canada and by the
Province of Ontario through the Ministry of Research \&
Innovation. P.~H. acknowledges support from ICFOstepstone,
funded by the Marie Sklodowska-Curie Co-funding of regional,
national and international programmes (GA665884) of the European
Commission, as well as by the Severo Ochoa 2016{--}2019' program at
ICFO (SEV{--}2015{--}0522), funded by the Spanish Ministry of Economy,
 Industry, and Competitiveness (MINECO).

\appendix

\section{Glossary}\label{Glossary}
This section contains an overview of terms discussed in the document which are
relevant for RBMs. For more detail we refer the reader to the code documentation
on \url{https://qucumber.readthedocs.io/en/stable/}, and
References~\cite{hinton_training_2002, hinton2012practical}.

\begin{itemize}

\item \textit{Batch}: A subset of data upon which the gradient is computed and
the network parameters are adjusted accordingly. A smaller batch size often
results in a more stochastic trajectory, while a large batch size more closely
approximates the exact gradient and has less variance.

\item \textit{Biases}: Adjustable parameters in an RBM, denoted by $b_j$ and
$c_i$ in Eq.~\eqref{RBMenergy}.

\item \textit{Contrastive divergence}: An approximate maximum-likelihood
learning algorithm for RBMs~\cite{hinton_training_2002}. CD estimates the
gradient of the effective energy~\eqref{RBMenergy} with respect to model
parameters by using Gibbs sampling to compare the generated and target
distributions.

\item \textit{Energy}: The energy of the joint configuration $(\bm{v}, \bm{h})$
of a RBM is defined as follows:
\begin{equation}
   E_{\bm{\lambda}}(\bm{v},\bm{h}) = - \sum\limits_{j=1}^{n_v} b_j v_j - \sum\limits_{i=1}^{n_h} c_i h_i - \sum\limits_{ij} h_i W_{ij} v_j. \label{RBMenergy}
\end{equation}

\item \textit{Effective energy}: Obtained from the energy by tracing out the
hidden units $\bm{h}$; often called the ``free energy'' in machine learning
literature.
\begin{equation}
   \mathcal{E}_{\bm{\lambda}}(\bm{v}) = - \sum\limits_{j=1}^{n_v} b_j v_j - \sum\limits_{i=1}^{n_h} \log \left[ 1 + \exp \left( \sum\limits_{j}^{n_v} W_{ij}v_j +c_i\right) \right]. \label{RBMeffectiveenergy}
\end{equation}

\item \textit{Epoch}: A single pass through an entire training set. For example,
with a training set of 1,000 samples and a batch size of 100, one epoch consists
of 10 updates of the parameters.

\item \textit{Gibbs sampling}: A Monte Carlo algorithm that samples from the conditional distribution of one variable, given the state of other variables.
In an RBM, the restricted weight connectivity allows Gibbs sampling between the visible ``block'', conditioned on the hidden ``block'', and vice versa.

\item \textit{Hidden units}: There are $n_h$ units in the hidden layer of the
RBM, denoted by the vector $\bm{h}=(h_1, \ldots, h_{n_h})$. The number of
hidden units can be adjusted to increase the representational capacity of
the RBM.\@

\item \textit{Hyperparameters:} A set of parameters that are not adjusted by a
neural network during training. Examples include the learning rate, number of
hidden units, batch size, and number of training epochs.

\item \textit{Joint distribution}: The RBM assigns a probability to each joint
configuration $(\bm v, \bm h)$ according to the Boltzmann distribution:
\begin{equation}
   p_{\bm{\lambda}}(\bm{v},\bm{h}) = \frac{1}{Z_{\bm{\lambda}}} e^{-E_{\bm{\lambda}}(\bm{v},\bm{h})}.
\end{equation}

\item \textit{KL divergence}: The Kullback-Leibler divergence, or relative
entropy, is a measure of the ``distance'' between two probability distributions
$P$ and $Q$, defined as:
\begin{equation}
\label{Eq.KLdiv}
\mathrm{KL}(P\:||\:Q)=\sum_{\bm{v}}P(\bm{v})\log\frac{P(\bm{v})}{Q(\bm{v})} .
\end{equation}
The KL divergence between two identical distributions is zero.
Note that it is not symmetric between $P$ and $Q$.

\item \textit{Learning rate}: The step size used in the gradient descent
algorithm for the optimization of the network parameters. A small learning rate
may result in better optimization but will take more time to converge. If the
learning rate is too high, training might not converge or will find a poor optimum.

\item \textit{Marginal distribution}: Obtained by marginalizing out the hidden layer from the joint distribution via
\begin{equation}\label{Eq:marginal_distribution}
   p_{\bm{\lambda}}(\bm{v}) = \frac{1}{Z_{\bm{\lambda}}} \sum_{\bm{h}} e^{-E_{\bm{\lambda}}(\bm{v},\bm{h})} = \frac{1}{Z_{\bm{\lambda}}} e^{- \mathcal{E}_{\bm{\lambda}}(\bm{v})}.
\end{equation}

\item \textit{QuCumber}: A quantum calculator used for many-body eigenstate reconstruction.

\item \textit{Parameters}: The set of weights and biases $\bm{\lambda} = \{\bm{W},\bm{b},\bm{c}\}$ characterizing the RBM energy function. These are adjusted during training.

\item \textit{Partition function}: The normalizing constant of the Boltzmann
distribution. It is obtained by tracing over all possible pairs of visible and
hidden vectors:
\begin{equation}
   Z_{\bm{\lambda}} = \sum\limits_{\bm{v},\bm{h}}e^{-E_{\bm{\lambda}}(\bm{v},\bm{h})}.
\end{equation}

\item \textit{Restricted Boltzmann Machine}: A two-layer network with
bidirectionally connected stochastic processing units. ``Restricted'' refers to
the connections (or weights) between the visible and hidden units: each visible
unit is connected with each hidden unit, but there are no intra-layer connections.

\item \textit{Visible units}: There are $n_v$ units in the visible layer of the
RBM, denoted by the vector $\bm{v}=(v_1,\dots,v_{n_v})$. These units correspond
to the physical degrees of freedom. In the cases considered in this paper, the
number of visible units is equal to the number of spins $N$.

\item \textit{Weights}: $W_{ij}$ is the symmetric connection or interaction
between the visible unit $v_j$ and the hidden unit $h_i$.

\end{itemize}

\bibliography{bib/RBMs,bib/RGandDL,bib/TNandDL,bib/MLtomography,bib/RBMphysics,bib/bibliography,bib/representationsDL,bib/moreReferences}
\nolinenumbers{}
\end{document}